# A good diagram is valuable despite the choice of a mathematical approach to problem solving

Alexandru Maries and Chandralekha Singh

*Department of Physics and Astronomy, University of Pittsburgh, Pittsburgh, PA 15260*

**Abstract.** Drawing appropriate diagrams is a useful problem solving heuristic that can transform a problem into a representation that is easier to exploit for solving the problem. A major focus while helping introductory physics students learn problem solving is to help them appreciate that drawing diagrams facilitates problem solution. We conducted an investigation in which 118 students in an algebra-based introductory physics course were subjected to two different interventions during the problem solving in recitation quizzes throughout the semester. Here, we discuss the problem solving performance of students in different intervention groups for two problems involving standing waves in tubes, one which was given in a quiz and the other in a midterm exam. These problems can be solved using two different methods, one involving a diagrammatic representation and the other involving mostly mathematical manipulation of equations. In the quiz, students were either (1) asked to solve the problem in which a partial diagram was provided or (2) explicitly asked to draw a diagram. A comparison group was not given any instruction regarding diagrams. Students in group (1), who were given the partial diagram, could not use that partial diagram by itself to solve the problem. The partial diagram was simply intended as a hint for students to complete the diagram and follow the diagrammatic approach. However, we find an opposite effect, namely, that students given this diagram were less likely to draw productive diagrams and performed worse than students in the other groups. Moreover, we find that students who drew a productive diagram performed better than those who did not draw a productive diagram even if they primarily used a mathematical approach. Interviews with individual students who were asked to solve the problem provided further insight.

**Keywords:** diagrammatic representations, diagrams, problem solving, physics education research

**PACS:** 01.40Fk,01.40.gb,01.40G-

## INTRODUCTION

Drawing diagrams is a useful problem solving heuristic. Diagrammatic representations have been shown to be superior to verbal representations when solving problems [1]. This is one of the reasons why physics experts automatically employ diagrams while solving problems. However, introductory physics students need explicit help understanding that drawing a diagram is an important step in organizing and simplifying the information given into a representation more suitable to further analysis including mathematical manipulation. Previous research shows that students who draw diagrams, even if they are not rewarded for it, are more successful problem solvers [2]. Here we describe a study involving problem solving in an introductory algebra-based course and discuss how students' performance is influenced by drawing productive diagrams even if they chose a primarily mathematical approach to problem solving. We also describe how prompting students to draw a diagram vs. providing a partial diagram impacts their performance.

## METHODOLOGY

A class of 118 introductory physics students (algebra-based course) was broken up into three different recitations. All recitations were taught in the traditional way in which the teaching assistant (TA) worked out problems similar to the homework problems and then gave a 15-20 minute quiz at the end of the class. Students in all recitations attended the same lectures, were assigned the same homework, and had the same exams and quizzes. In the recitation quizzes throughout the semester, the three groups were given the same problems but with the following interventions: in each quiz problem, the first intervention group, which we refer to as the "prompt only group" or "PO", was given an explicit prompt to draw a diagram along with the problem statement. The second intervention group (referred to as the "diagram only group" or "DO") was given a diagram drawn by the instructor that was meant to aid in solving the problem and the third group was the comparison group and was not given any diagram or explicit instruction to draw a diagram with the problem statement ("no support group" or "NS").

The sizes of the different recitation groups varied from 22 to 55 students because the students were not assigned a particular recitation, they could go to whichever recitation they wanted. For the same reason, the sizes of each recitation group also varied from week to week, although not as drastically because most students (≈80%) would stick with a particular recitation. Furthermore, each intervention was not matched to a particular recitation. For example, in one week, NS was the Tuesday recitation while in another week NS was a different recitation section. This is important because it implies that individual students were subjected to different interventions from week to week and we therefore do not expect cumulative effects due to the same group of students always being subjected to the same intervention.

In order to ensure homogeneity of grading, we developed a rubric for each problem analyzed and made sure that there was at least 90% inter-rater-reliability between two different raters. The development of the rubric for each problem went through an iterative process. During the development of the rubric, the two raters discussed some students' scores separately from those obtained using the preliminary version of the rubric and adjusted the rubric if it was agreed that the version of the rubric was too stringent or too generous. After each adjustment to the rubric, all the students were graded again on the improved rubric.

Here, we discuss two similar problems which involve standing waves in tubes. One was given in a quiz in which the interventions discussed above were implemented and the other was given in a midterm exam in which all students received the same instructions. The problems are as follows:

**Quiz problem:**
"A tube with air is open at only one end and has a length of 1.5 m. This tube sustains a standing wave at its third harmonic. What is the distance between a node and the adjacent antinode?"

**Midterm exam problem:**
The midterm problem was identical to the quiz problem except that the tube was open at both ends instead of just one.

There are two approaches to solving the quiz problem (the midterm exam problem can also be solved by employing a very similar strategy for a tube that is open at both ends). One strategy is to draw the standing wave pattern as shown in Figure 1.

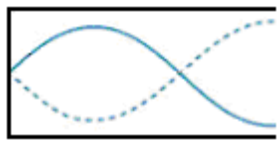

**FIGURE 1:** Third harmonic of a standing wave in a tube open at only one end.

Then, for example, one can identify that three node to antinode distances fit in the tube with length $L$=1.5 m. Therefore, the distance between a node and the adjacent antinode is 1.5/3 = 0.5 m. This diagrammatic approach is a more expert-like approach because it requires understanding of a physics concept in its diagrammatic representation (third harmonic of a standing wave) and how it applies to a tube which is open at only one end (displacement node at the closed end and antinode at the open end). The second approach to solving this problem is to use the equation for the frequency of the $n^{th}$ harmonic of a standing wave in a tube of length $L$ open at one end, $f_n = \frac{nv}{4L}$, which was provided to students, and the relation between the speed $v$, frequency $f$ and wavelength of a wave, $v = f \lambda$ (not provided)**,** solve for the wavelength $\lambda$ given $L$ and $n$ and finally divide the wavelength obtained by 4 to get the distance between a node and the adjacent antinode. We refer to this latter approach as the "mathematical" approach because it does not necessarily require understanding the physics principles involved, up to finding the wavelength and the two equations can be used as mathematical algorithms if students have the mathematical skills required to manipulate them.

The "partial" diagram given to the students in intervention DO contained an empty tube. The rationale for giving the empty tube was to investigate if the empty tube aids in problem solving and whether students in this intervention are more likely to draw the standing wave and follow a more expert-like approach.

The quiz problem was also given to 26 first year physics graduate students (physics experts for this study) enrolled in a TA training course to benchmark the performance of introductory students and assess how often experts use the diagrammatic approach (which was hypothesized to be a more expert-like approach). It is a straightforward exercise for a physics graduate student to solve for the wavelength using the mathematical approach discussed earlier. However, we found that 76% of them elected to draw a diagram to solve the problem and completely ignored the equations provided to them, thus confirming our hypothesis that experts are more likely to follow the diagrammatic than the mathematical approach to solve this problem.

We investigated how the different interventions impacted the students in terms of how likely they were to draw productive diagrams. How much value one derives from drawing a particular type of diagram and how the person employs the diagram (and the process of drawing it) to solve a problem depend on the expertise of the individual. However, for the purposes of this research, a diagram was considered to be productive if it could have aided students in solving the problem based upon a cognitive task analysis of the problem. The productive diagrams were classified in two broad categories: diagrams of third harmonics of waves in tubes (whether correct or incorrect) and diagrams of one wavelength (drawn as a single sinusoidal wave). An attempt to draw a third harmonic was considered to be productive even if it did not represent a third harmonic for the tube open at one end and closed at the other. This type of diagram can guide the problem solving process via the more expert-like approach. The second type of diagram (diagrams of one wavelength of a single sinusoidal wave) was considered to be productive because it could be used to determine what fraction of a wavelength is the distance between a node and the adjacent antinode.

Furthermore, because there are two approaches to the solution of this standing wave problem, one primarily diagrammatic and another primarily based on mathematical manipulations, rubrics were developed to score the performance of students employing each approach. Due to space constraints, we will only present the results for students who primarily employed the mathematical approach. The summary of the rubric used to score students who employed this approach is shown in Table 1.

**TABLE 1.** Summary of the rubric used to score the performance of students employing the mathematical approach out of 10 points (p).

| Correct Knowledge | |
|---|---|
| 1. Used given equation $f_n = nv/4L$ | 1 p |
| 2. Chose $n = 3$ | 1 p |
| 3. Wrote down $v = f\lambda$ | 3 p |
| 4. Solved for $\lambda$ correctly | 2 p |
| 5. Found answer by dividing $\lambda$ by 4 | 2 p |
| 6. Correct unit for answer | 1 p |
| **Incorrect Ideas** | |
| 1. Used incorrect equation (-1 p) | |
| 2. Chose value for $n$ other than 3 (-1 p) | |
| 3.1 Did not write $v = f\lambda$ (-3 p) | |
| 3.2 Tried to write down $v = f\lambda$, but made a mistake (e.g., $v = f/\lambda$ or similar) (-2 p) | |
| 4.1 Did not solve for $\lambda$ (-2 p) | |
| 4.2 Used a value for $v$ other than speed of sound (-1 p) | |
| 4.3 Obtained incorrect $\lambda$ (-1 p) | |
| 4.4 Unclear how $\lambda$ was found or other error (-1 p) | |
| 5. Did not divide $\lambda$ by 4 to obtain the answer or did not obtain an answer (-2 p) | |
| 6. Incorrect units (-1 p) | |

Table 1 shows that there are two parts to the rubric: Correct Knowledge and Incorrect Ideas. Table 1 also shows that in the Correct Knowledge part, the problem was divided into different sections and points were assigned to each section. Each student starts out with 10 points and in the Incorrect Ideas part, the common mistakes students made in each section and the number of points that were deducted for each of those mistakes are listed. It is important to note that each mistake is connected to a particular section (the mistakes labeled 1 and 2 are for the first and second sections, the two mistakes labeled 3.1 and 3.2 are for the third section and so on) and that for each section, the rubric cannot be used to subtract more points than that section is worth. For example, the two mistakes in section 3 (3.1 and 3.2) are mutually exclusive. Similarly, mistake 4.1 is exclusive with all other mistakes in section 4 and mistakes 4.3 and 4.4 are mutually exclusive. Finally, if the mistake a student made was not common and not in the rubric, it would correspond to the mistake labeled as 4.4.

In addition to analyzing the quantitative data, interviews were conducted with eight students using a think-aloud protocol [3] in order to obtain an in-depth account of their difficulties while solving the quiz problem and in addition provide some insights that would account for the performance of these students. Due to space constraints, the interviews will be discussed only briefly.

# RESULTS

The numbers of students, averages and standard deviations for the scores of students in the different intervention groups are shown in Table 2. *T*-tests on the data in Table 2 reveal that students in DO performed statistically significantly worse than students in PO and NS (p=0.007 for PO-DO comparison and p<0.001 for DO-NS comparison). The difference between PO and NS is not statistically significant. In the midterm exam problem, all groups exhibited comparable performance (no statistically significant differences).

**TABLE 2.** Numbers (N), averages (Avg.) and standard deviations (St. dev.) for the scores of students in the different intervention groups in the quiz problem.

| Quiz | N | Avg. | St. dev. |
|---|---|---|---|
| PO | 50 | 8.0 | 1.7 |
| DO | 39 | 6.6 | 2.3 |
| NS | 29 | 8.5 | 1.1 |

Table 3 shows the percentages of students who drew productive diagrams (either diagrams of third harmonics of standing waves, whether correct or not, or diagrams of one wavelength of a single sinusoidal wave) from each intervention group and Table 4 shows the p values for comparing these percentages for different intervention groups. It appears that students in PO are more likely than students in DO and NS to draw productive diagrams. Also, it appears that students in DO were somewhat less likely than students in the comparison group (NS) to draw productive diagrams.

**TABLE 3.** Percentages of students who drew productive diagrams in the quiz for each intervention group (PO, DO and NS)

| Quiz | PO | DO | NS |
|---|---|---|---|
| Percent of students who drew productive diagram | 96% | 60% | 79% |

**TABLE 4.** p values for chi-square tests comparing the different groups (PO, DO, NS) in terms of percentage of students who drew productive diagrams in the quiz.

| Quiz | PO-DO | PO-NS | DO-NS |
|---|---|---|---|
| p value | <0.001 | 0.046[a] | 0.076 |

a. Fisher's exact test was used instead of chi-square test because some expected cell frequencies were smaller than 10, see.[5].

We investigated how drawing a diagram impacted the scores of students who used a primarily the mathematical approach in the quiz problem and in the midterm exam problem. Table 5 shows the numbers, averages and standard deviations of students who primarily employed a mathematical approach, but also drew a productive diagram and students who primarily employed a mathematical approach but did not draw a productive diagram. Both in the midterm exam and quiz, the students who primarily used mathematical approach to solve the problems but also drew a productive diagram, performed better than those who primarily used mathematical approach but did not draw a productive diagram (p = 0.002 for the quiz and p = 0.006 for the midterm exam).

**TABLE 5.** Numbers (N), averages (Avg.) and standard deviations (St. dev.) for the scores of students who primarily used math, but also drew a productive diagram, and students who used math without a productive diagram both in the quiz and midterm.

| Quiz | N | Avg. | St. dev. |
|---|---|---|---|
| Primarily used math, but also drew a productive diagram | 45 | 8.1 | 1.7 |
| Used math without a productive diagram | 24 | 6.6 | 2.1 |
| **Midterm** | **N** | **Avg.** | **St. dev.** |
| Primarily used math, but also drew a productive diagram | 68 | 8.8 | 2.0 |
| Used math without a productive diagram | 24 | 7.3 | 2.3 |

## DISCUSSION AND SUMMARY

We find that students who were given a diagram of an empty tube performed statistically worse than those who were asked to draw a diagram as well as those who were not given any instructions regarding diagrams. We find a similar result while investigating introductory students' performance on two problems in electrostatics that involved considerations of initial and final situations [4]. The research in Ref. [4] involved the same methodology as described here. However, the diagrams given to students in group DO were very similar to what most instructors would draw in order to solve those problems and they were intended as scaffolding support. The given diagrams had the opposite effect, apparently worsening their performance as compared to students in the other two groups. In the research presented here, the diagram given to students in group DO was not a fully functional one. Rather, the empty tube provided was intended as a hint or prompt to get students to attempt to solve the problem in an expert-like manner (drawing a diagram of the third harmonic of a standing wave and using it to solve the problem). However, similar to the study involving electrostatics, here, we also find that providing the diagram of the empty tube had the opposite effect from what was intended. In particular, the students who were given this diagram drew fewer productive diagrams than those who were not provided a diagram. This may partly account for the diminished performance of students provided with a diagram.

We also find that among the students who employed a primarily mathematical approach, those who drew productive diagrams performed better than those who did not. Thus, it appears that a diagram and the process of drawing a diagram (which is part of the conceptual analysis of the problem) helps improve students' scores even if their chosen approach to proceed further with the problem solving process does not explicitly require a diagram. In the problems discussed here, among the students who primarily chose the mathematical approach but also drew diagrams, both in the quiz and in the midterm, about 80% of them attempted to draw diagrams of the third harmonic. Interviews suggest that students in a similar situation were trying to make sense of the problem conceptually even though drawing a diagram of a third harmonic does not necessarily help one solve the problem if the chosen approach is the mathematical one. This sense-making may partly account for the improved performance of students who used the mathematical approach to solve the problem but also drew productive diagrams as compared to those who did not draw productive diagrams and only used the mathematical approach. The think-aloud interviews also suggest that the students who explicitly used the productive diagrams they drew were less likely to make mistakes than the students who did not use them.